# The 2nd Generation z(Redshift) and Early Universe Spectrometer Part I: First-light observation of a highly lensed local-ULIRG analog at high-z

Short Title:
ZEUS-2 First Light Observations

Carl Ferkinhoff[1]
Drew Brisbin[1]
Stephen Parshley[1]
Thomas Nikola[1]
Gordon J. Stacey[1]
Justin Schoenwald[1]
James L. Higdon[2]
Sarah J. U. Higdon[2]
Aprajita Verma[3]
Dominik Riechers[1]
Steven Hailey-Dunsheath[4]
Karl M. Menten[5]
Rolf Güsten[5]
Axel Weiß[5]
Kent Irwin[6]
Hsiao M. Cho[6]
Michael Niemack[7]
Mark Halpern[8]
Mandana Amiri[8]
Matthew Hasselfield[9]
D.V. Wiebe[8]
Peter A. R. Ade[10]
Carol E. Tucker[10]

---

[1] Department of Astronomy, Cornell University, Ithaca, NY 14853, USA; carl.ferkinhoff@cornell.edu
[2] Department of Physics, Georgia Southern University, Statesboro, GA 30460, USA
[3] Department of Physics, University of Oxford, Oxford, OX1 3RH, UK
[4] California Institute of Technology, Pasadena, CA 91125, USA
[5] Max-Planck-Institut für Radioastronomie, Auf dem Hügel 69, 53121 Bonn, Germany
[6] NIST Boulder, Boulder, CO 80305, USA
[7] Department of Physics, Cornell University, Ithaca, NY 14853, USA
[8] Department of Physics and Astronomy, University of British Columbia, Vancouver, B.C., V6T 1Z1, Canada
[9] Department of Astrophysical Sciences, Princeton University, Princeton, NJ 08544, USA
[10] Department of Physics and Astronomy, Cardiff University, Cardiff CF24 3AA, UK

Abstract:

We report first science results from our new spectrometer, the 2$^{nd}$ generation z(Redshift) and Early Universe Spectrometer (ZEUS-2), recently commissioned on the Atacama Pathfinder Experiment telescope (APEX). ZEUS-2 is a submillimeter grating spectrometer optimized for detecting the faint and broad lines from distant galaxies that are redshifted into the telluric windows from 200 to 850 microns. It utilizes a focal plane array of transition-edge sensed bolometers, the first use of these arrays for astrophysical spectroscopy. ZEUS-2 promises to be an important tool for studying galaxies in the years to come due to its synergy with ALMA and its capabilities in the short submillimeter windows that are unique in the post Herschel era. Here we report on our first detection of the [CII] 158 µm line with ZEUS-2. We detect the line at z ~ 1.8 from H-ATLAS J091043.1-000322 with a line flux of $(6.44 \pm 0.42) \times 10^{-18}$ W m$^{-2}$. Combined with its far-infrared luminosity and a new Herschel-PACS detection of the [OI] 63 µm line we model the line emission as coming from a photo-dissociation region with far-ultraviolet radiation field, G ~ 2 x 10$^4$ G$_0$, gas density, n ~ 1 x 10$^3$ cm$^{-3}$ and size between ~ 0.4 and 1 kpc. Based on this model, we conclude that H-ATLAS J091043.1-000322 is a high redshift analogue of a local ultra-luminous infrared galaxy, i.e. it is likely the site of a compact starburst due to a major merger. Further identification of these merging systems is important for constraining galaxy formation and evolution models.

Subject Headings: galaxies: individual (H-ATLAS J091043.1-000322) galaxies: high-redshift – galaxies: starburst – galaxies: active – submillimeter: galaxies

1. Introduction

The submillimeter view of the early universe is rapidly evolving. Improvements in instrumentation capabilities over the past 15 years have revealed the submillimeter band (~200 to 850 μm) as a unique probe of star formation in the early universe. These studies were launched by small to modest-sized ground-based surveys, largely of one or two colors, and moderate angular resolution (~10") far exceeding the previous submillimeter surveys of cosmology experiments (e.g. Smail, Ivison & Blain 1997; Berger et al. 1998; Hughes et al. 1998; Weiss et al. 2009; Coppin et al. 2006). Herschel expanded these studies to include both larger scales and multiple colors. From the ground we began high-redshift spectroscopic surveys of far-infrared fine structure lines in the 350 and 450 μm telluric windows with our 1$^{st}$ generation z(Redshift) and Early Universe Spectrometer (ZEUS-1) on the Caltech Submillimeter Observatory ([CII] – Hailey-Dunsheath et al. 2010; Stacey et al. 2010; Brisbin et al. 2013, in prep.; [OIII] – Ferkinhoff et al. 2010; [NII] – Ferkinhoff et al. 201). The Herschel-SPIRE Fourier transform spectrometer enabled similar submillimeter spectroscopy that was un-obscured by telluric absorption. With the advent of the Atacama Large Millimeter Array (ALMA), studies at wavelengths between 0.43 and 3 millimeters are now making great leaps forward in terms of sensitivity and spatial resolution. However, as Herschel is no longer operating and ALMA has yet to open the 350 μm (Band 10) window, there are limited opportunities to detect far-infrared lines from high redshift systems at wavelengths shorter than ~ 420 μm. For the [CII] 158 μm line, typically the bright FIR fine-structure line, this means ALMA can observe galaxies no closer than z ~1.7 until band 10 is deployed.

Recently, we have commissioned our new instrument, the 2$^{nd}$ generation z(Redshift) and Early Universe Spectrometer (ZEUS-2) on the Atacama Pathfinder Experiment (APEX) in

Chile[11]. When fully deployed, ZEUS-2 will have wavelength coverage between 200 and 850 microns. For its first run, reported here, ZEUS-2 only accessed the 350 and 450 μm telluric windows. With a spectral resolving power of ~1000 and background limited detectors, ZEUS-2 is optimized for detecting broad lines from galaxies (line-widths of ~200 – 300 km/s). Within these windows ZEUS-2 can rapidly survey high-redshift galaxies, detecting far-IR lines that are spatially unresolved in a typical ZEUS-2/APEX beam (~8 arcsec) yet suitable for follow-up at high spatial resolution with ALMA. In local systems ZEUS-2 can simultaneously map their emission in the $^{13}$CO 6-5, CO 7-6, [NII] 205 μm, [CI] 370 μm and the [CI] 605 μm lines that provide important information about the ionized, atomic, and molecular phases of the ISM. Part II of this paper (Ferkinhoff et al. 2013, in prep) discusses the instrument design and performance in detail.

The star formation and ISM of high-z systems can be studied with ZEUS-2 through the FIR fine-structure lines of carbon, oxygen, nitrogen, and their various ions that are redshifted into the submillimeter atmospheric windows. These lines characterize both the physical conditions of the gas and the ambient radiation fields. They are especially useful for studying dusty galaxies in the early universe where the commonly used optical emission line tracers of the ISM (e.g. [OII] 3726, 3729 Å and [OIII] 4959, 5007 Å) undergo significant extinction, thereby limiting their use as astrophysical probes. The FIR fine structure lines have been extensively studied in nearby extragalactic systems with the Kuiper Airborne Observatory (KAO), the Infrared Space Observatory (ISO) and recently with Herschel (e.g. Crawford et al. 1986, Stacey et al. 1991, Lord et al. 1996, Luhman et al. 1998, Malhotra et al. 2001, Negishi et al. 2001, Brauher et al. 2008, Gracia-Carpio et al. 2011). These studies were extended to high redshift galaxies with Herschel (cf. Ivison et al. 2010, Sturm et al. 2010, Valtchanov et al. 2011, Coppin et al. 2012, Verma et al. 2013 in prep) and regularly detected from the ground with ZEUS-1, as mentioned above, as well as with telescopes like the IRAM 30 meter, the Plateau de Bure Interferometer (PdBI), and ALMA (e.g. Maiolino et al. 2005, 2009, Swinbank et al. 2012, DeBreuck et al. 2011, Cox et al. 2011, DeCarli et al. 2012, Carilli et al. 2013, Wang et al. 2013). Most recently, ALMA and the PdBI are now enabling the spatially resolved studies of these important cooling lines (Wagg et al. 2012, Galhervani et al. 2012).

---

[11] This publication is based on data acquired with the Atacama Pathfinder Experiment (APEX). APEX is a collaboration between the Max-Planck-Institut fur Radioastronomie, the European Southern Observatory, and the Onsala Space Observatory.

1.1 Studies of far-infrared [CII] and [OI] emission

In nearby systems the [CII] 158 µm and [OI] 63 µm lines are typically the brightest of the FIR lines, with line luminosities between ~0.1 and 1 percent of the FIR luminosity (e.g., Crawford et al. 1986, Stacey et al. 1991, Malhotra et al. 2001). These lines arise in warm and dense photo-dissociation regions (PDRs) that are formed on the surfaces of molecular clouds (Tielens and Hollenbach, 1985) by far-ultraviolet radiation (FUV; 6 eV < $h\nu$ < 13.6 eV) from early-type stars. The gas in PDRs is heated by electrons photo-ejected off small dust-grains by the FUV field and cooled by the [CII] and [OI] lines. The relative strength of the lines depends on both the gas density and strength of the FUV field where the [OI] 63 µm line dominates the cooling at higher densities and FUV fields. Over the typical range of PDR gas densities for extra-galactic sources (n ~ $10^3$ to $10^5$ cm$^{-3}$), the [CII]/FIR ratio is inversely proportional to the strength of the FUV field, so that it indicates the strength of the FUV field and source size (Stacey et al. 2010).

ISO studies found a decline in the [CII]/FIR ratio for increasingly luminous, nearby galaxies (Mahotra et al. 2001, Luhman et al. 2003). The decline is attributed to dust grains acquiring greater positive charge due to photo-ejection of electrons in the extreme FUV environment of the compact and intense star forming regions that are induced by major mergers. The accumulated grain charge reduces the gas heating efficiency of the ejected electron limiting the collisional excitation rate of ionized carbon and tempering the rise in [CII] line emission. As a result, in clouds with densities near or above the critical density of the transition (~ $2 \times 10^3$ cm$^{-3}$) that are also exposed to high FUV fields, the excitation of the [CII] line emitting level effectively saturates and higher gas densities and temperatures do not lead to a larger population in the excited level. The column density of $C^+$ in such environments is determined by dust extinction of the UV photons capable of ionizing carbon, so that the $C^+$ column only grows logarithmically with FUV flux. Meanwhile the FIR emission continues to increase linearly for increasing FUV fields due to the dust reprocessing of the UV photons (Stacey et al. 2010). This explanation is also supported by the suppression of all the FIR fine-structures lines in the FIR luminous systems seen in the study by Gracia-Carpio et al. (2011). In some ULIRGs, however, there may be a significant AGN contribution to the FIR luminosity significantly reducing the [CII]/FIR line ratio if the AGN contribution to the FIR luminosity is not accounted for (Sargsyan et al. 2012).

The first [CII] survey of high-z, FIR luminous systems ($L_{FIR} > 10^{12.5}$ $L_\odot$) showed that star formation dominated galaxies in the early universe did not have the low [CII]/FIR ratios characteristic of the ultra-luminous systems nearby (Stacey et al. 2010). Surprisingly, their ratios were similar to local moderate-luminosity systems indicating kpc-scale and moderate- intensity star formation – not the concentrated and merger-induced starbursts seen in local ULIRGs. The high-z systems with low [CII]/FIR ratios that are similar to the ratios of local ULIRGs, have buried AGN that dominate their energetics (Stacey et al. 2010). In this work we report on observations of a lensed ULIRG that we conclude is a high-z analogue of local ULIRGs: a system with strong FUV fields and compact, sub-kiloparsec, star formation induced by a major merger.

1.2  This Paper: First ZEUS-2 detection of a spectral line from a high redshift galaxy

In this paper we report on the first light spectra of ZEUS-2 on APEX obtained in November 2012. We strongly (~ 12σ) detect the [CII] 158 micron line from H-ATLAS J091043.1-000322 in ~ 66 minutes of integration time. For our analysis we combine our [CII] detection with a recent Herschel detection of the [OI] 63 μm line (Verma et al. 2013, in prep), previous CO (Lupu et al. 2012), and archival Hubble Space Telescope (HST) images. This work is the first spectroscopic use of a transition edged sensed bolometer array, the state of the art in submillimeter detector technology, and demonstrates the scientific potential of ZEUS-2.

In Section 2 of this paper we describe the source, the ZEUS-2 observations, data reduction and a gravitational lensing model based on archival HST Images. Section 3 presents a combined analysis of the new spectroscopic measurements. The implications of this analysis are discussed in Section 4. Lastly we give our concluding remarks in Section 5. Throughout this paper we have adopted the cosmological parameters of $\Omega_\Lambda = 0.73$, $\Omega_m = 0.27$ and $H_0 = 71$ km s$^{-1}$ Mpc$^{-1}$ (Spergel et al. 2003). We define the far-infrared luminosities to be from 42.5 to 122.5 μm (covering the IRAS 60 and 100 μm bands) following the prescription of Helou et al. (1986). Some authors extend the FIR luminosity to include wavelengths up to 500 microns; these values are typically ~1.5 times larger than the luminosities we report here. The total infrared luminosity, $L_{IR}$, is the integrated luminosity between 8 and 1000 microns and is ~2 times larger than the FIR luminosity as we define above.

2.  The Source, Observations, and Lensing Model

2.1 H-ATLAS J091043.1-000322

First reported by Negrello et al. (2010), H-ATLAS J091043.1-000322 (hereafter SDP11) was identified in the Herschel guaranteed time program Herschel Astrophysical Terahertz Large Area Survey (H-ATLAS, Eales et al. 2010). Subsequent spectroscopic follow-up with Z-spec on the CSO identified several mid-J CO transitions at $z = 1.786$ (Lupu et al. 2012). Based on its exceptionally bright line fluxes, its very large apparent FIR luminosity[12] ($L_{FIR} = 3.88 \times 10^{13}$ $L_\odot$) and the identification of a foreground lensing galaxy at $z = 0.793$, Negrello et al. conclude that SDP11 is a highly gravitationally lensed system, with more moderate intrinsic luminosity. Negrello et al. do not however, constrain the magnification. Using archival HST observations and a simple lensing model we confirm that SDP11 is indeed a lensed galaxy with magnification, μ, between 7 and 29, see (Section 2.3 below), so that the intrinsic molecular-gas mass and far-IR luminosity are ~ $(0.8 – 3.4) \times 10^{10}$ $M_\odot$ and $L_{FIR} = (1.3 - 5.4) \times 10^{12}$ $L_\odot$, respectively. Table 1 lists the observed source properties from this work and the literature.

2.2 Observations

Using ZEUS-2 on APEX we observed SDP11 on 2012 November 17 under very good observing conditions (0.48 mm of precipitable water vapor). Spectra were obtained in standard chopping and beam switching mode with a chopper throw of 60". We obtained five, 13.2 minute integrations for a total integration time of 66 minutes. The ZEUS-2 velocity resolution at the observed wavelength (439.6 μm) is ~ 340 km/s. The ZEUS-2/APEX beam is 8" (~68 kpc at $z = 1.78$), as measured on Uranus, which fully contains SDP11. Uranus also served as our flux calibrator using a brightness temperature of 68 K at 450 microns as reported by Hildebrand et al.

---

[12] Negrello et al. 2010 and Lupu et al. 2012 report different values for the FIR luminosities of SDP11 likely due to different definitions of the FIR luminosities. To ensure what we are consistent with our ZEUS-1 sample in Brisbin et al 2013 (in prep.) we have taken the photometry reported in Negrello et al. and performed our own fit of the spectral energy distribution (SED) using the SED templates of Siebenmorgen & Krugel (2004) and integrating the SED according to our definition of the FIR luminosity in Section 1.2.

(1985). Figure 1 shows the detected [CII] 158 micron line with a flux of $(6.44 \pm 0.42) \times 10^{-18}$ W m$^{-2}$.

2.3 Lensing Model

Figure 2 shows a Hubble Wide Field Camera 3 (WFC3) near infrared image (F110W) of SDP11, centered on the lensing galaxy SDSS J091043-000323 (hereafter SDSSJ0910) at z = 0.793 (central elliptical at position 0", 0"). The contours show the HST/WFC3 F160W image divided by the F110W image, i.e. the 1.5 micron/1.1 micron image[13]. Based on the SED modeling of Negrello et al. (2010) who separately fit an SED to SDP11 and the lensing galaxy SDSSJ0910, we expect both SDSSJ0910 and SDP11 to be brighter in the 1.5 micron band. However the 1.5/1.1 micron ratio for SDP11 is ~4 times larger than that of SDSSJ0910 and the divided image should show a lensed imaged of SDP11. Indeed, the divided image (i.e. the contours if Figure 2) clearly show an Einstein ring centered on SDSSJ0910. To create a lens model we take the positions of the ring's peak values relative to the centroid of SDSSJ0910 to correspond to lensed images of SDP11 assuming it is a point source. We input the position and values of the peaks in the divided image into LENSMODEL (Keeton 2001) in order to model the gravitational lens system and reproduce the peaks of the Einstein ring. The squares indicate the peak positions while the triangles indicate the predicted image locations based on our model solution. While we obtain an excellent model fit to the image positions and fluxes ($\tilde{\chi}^2 \sim 1$), to determine the magnification and reproduce the observed Einstein ring we use the point-source derived model of the lens, but replace the point source with an extended source. Adding an extended source adds uncertainty to the magnification factor since we do not know how well the width of the ring in the divided image represents the true source size. With that caveat in mind, a source with a Gaussian light profile and half-light radius of 2.1 ± 1.3 kpc is able to reproduce the observed width of the Einstein ring in Figure 2 and results in a flux magnification, µ, of 18 ± 11. This is of course the magnification of the NIR emission in SDP11. The magnification of SDP11's submillimeter emission could be different if it is distributed differently relative to the position of the lens. Given the lack of high resolution submillimeter images we adopt the NIR magnification above throughout this paper, with of course the caveat in mind that the submillimeter magnification may in fact be different.

---

[13] These images were obtained from the Hubble Legacy Archive and where part of HST Cycle 18 proposal 12194 and are described in detail in Negrello et al. (in prep.) and Dye et al. (in prep.).

3. Analysis

3.1 The [CII] 158 µm to FIR Luminosity Ratio: Intense FUV Fields

SDP11 has an apparent FIR luminosity of $3.88 \times 10^{13}$ $L_\odot$, placing it firmly in the hyper-luminous infrared galaxy class (HyLIRG, $L_{FIR} \geq 10^3 L_\odot$). Figure 3 shows a plot of $L_{[CII]}/L_{FIR}$ ratio, R, for SDP11 versus its apparent FIR luminosity along with samples of local galaxies—normal star forming galaxies and local ULIRGs—and high-z galaxies including the sources from Stacey et al. 2010 and Hailey-Dunsheath et al. 2010. The ratio R~(1.0 ± 0.3) × $10^{-3}$ for SDP11 falls between the average line ratios observed for star formation dominated systems ($L_{[CII]}/L_{FIR}$~3 × $10^{-3}$) and AGN dominated systems ($L_{[CII]}/L_{FIR}$~4 × $10^{-4}$) at high-z. Stacey et al. 2010 and Haley-Dunsheath et al. 2010 show that stronger UV fields may lower the value of R and for a given value of R, i.e. for a given FUV field, the FIR luminosity indicates the spatial scale of the emission. The ratio of SDP11, when compared to the Stacey et al. sample of sources with similar ratios, suggests a FUV field, G, of ~ 10000 $G_0$, where $G_0$ is the Habing field—the strength of the local interstellar radiation field corresponding to a flux of $1.6 \times 10^{-3}$ erg cm$^{-2}$ s$^{-1}$. Given the value of R from SDP11, its FIR luminosity suggests a source size greater than a kilo-parsec. However, once we account for the magnification of SDP11 its intrinsic FIR luminosity becomes similar to local ULIRGS as well. Magnification does not affect the estimate of the FUV field (G~10000 $G_o$), which is more intense than local normal galaxies or high-z star formation dominated systems, but the lower FIR luminosity now suggests sub-kiloparsec size emission like that found in nearby ULIRGs.

3.2 Modeling the Line and Continuum Emission

Comparisons between our [CII] line, the [OI] line, and the FIR continuum will provide tighter constraints on the ambient FUV fields and gas density in SDP11 as well as the nature of the source. Using the Herschel-PACS spectrometer on the Herschel Space Telescope, Verma et al. (2013, in prep) report a [OI] 63 µm line flux of $(7.5 \pm 2.3) \times 10^{-18}$ W/m$^2$. Typically the [CII] and [OI] lines are produced in PDRs associated with star forming regions. However, it is possible that both of these lines arise within the x-ray dominated regions (XDRs) produced in molecular clouds enveloping an AGN. Meijerink, Spaans, and Israel (2006) produce grids of observed flux for various FIR fine structure lines, including [CII] and [OI], across typical densities and x-ray fluxes of XDRs. The observed [OI]/[CII] ratio is $1.2 \pm 0.5$. This is only consistent with the lowest x-ray fluxes and densities ($F_x \sim 0.1$-$0.3$ erg cm$^{-2}$ s$^{-1}$ and n ~ 100 – 1000), so we expect that the line emission from SDP11 is likely not dominated by an XDR region associated with an AGN. Furthermore, since the currently available photometry of SDP11 can be fit solely by a star formation dominated spectral energy distribution (see Lupu et al. 2012), we proceed with an analysis of the observed [CII] and [OI] lines and far-IR continuum emission within a star formation dominated, PDR paradigm.

Kaufman et al. (2006) produce models grids of PDR regions over a range of densities and FUV fields which we use for our modeling. In performing the PDR modeling we assumed ~ 70% of the observed [CII] flux arise in the PDR, with most of the remaining 30% arising from the warm ionized medium (Oberst et al. 2008). The models of Kaufman et al. assume a single face-on cloud. In reality the PDRs in SDP11 are more complex than used in Kaufman et al., and if we assume they are spherical then we will detect [CII] and FIR emission from both the front and the back of the cloud, while we only see [OI] from the front because it is optically thick. To make use of the Kaufman et al. models we must then multiply our observed [OI] line flux by two to account for the emission that can be self-absorbed along the line of sight by intervening atomic oxygen. Both of these corrections are appropriate for analysis within a star formation paradigm (Kaufman et al. 1999). In figure 4 we plot the corrected [CII]/FIR and ([CII] + [OI])/FIR flux ratios for SDP11. This gives a PDR model solution of G = 20,000 $G_0$ and a gas density n = 2500 $cm^{-3}$. In principle, CO observations are also useful in constraining the PDR model. The observed CO lines are in agreement with our [CII], [OI], and FIR-constrained model, however due to the low signal-to-noise detection of these lines from SDP11, their inclusion does not help to constrain it further as is evident by Fig. 4.

Our derived FUV field and gas density are consistent with values observed in both local ULIRGS (e.g. Luhman et al. 2003) and high-z sources dominated by AGN (e.g. Stacey et al. 2010) agreeing with our conclusions above that were drawn from the [CII] to continuum ratio alone. The observations of the [CII], [OI], and FIR alone are not enough to unambiguously identify the nature of the source. However, the observed [CII], [OI], and FIR luminosity can be described entirely within a star formation paradigm. SDP11 most likely features a compact starburst similar to those in local ULIRGs. While this does not exclude the presence of an AGN, if one is present in SDP11 it likely does not contribute significantly to our observations or conclusions.

Having constrained the FUV field within SDP11 we estimate the size of the emission region, i.e. the size of the starburst, following the method from Stacey et al. (2010). This will provide additional clues to the nature of the source. Since the FIR emission arises from processing of the FUV radiation by dust, the ratio of the FIR luminosity to the FUV radiation field determines the size of the emitting regions. From Wolfire et al. (1990) the source diameter, D, is proportional to $(\lambda L_{FIR}/G)^{1/3}$ if the mean-free path ($\lambda$) of a FUV photon is small. If instead the mean-free path of a FUV photon is large then $D \propto (L_{FIR}/G)^{1/2}$. The constant of proportionality of these relations is determined by assuming that the mean-free path in SDP11 is the same as for M82, which has D ~ 300 pc (Joy et al. 1987), G ~ 1000 $G_0$ (Lord et al. 1996), and $L_{FIR}$ ~ 2.8 × $10^{10}$ $L_\odot$ (the average of the values reported in Colbert et al. 1999 and Rice et al. 1988). The source diameter is then between 0.4 and 0.9 kpc if we assume that SDP11 is lensed a factor between 7 and 29, and we account for the extremes in mean-free path as described above. This agrees with the size of the dust emission determined in Lupu et al. (2012) who estimate the solid angle, $\mu\Omega_d$, of the dust emission region in SDP11 to be 0.43 arcsec$^2$. Accounting for the magnification by gravitational lensing, this corresponds to a circular area with diameter between 0.6 and 1.36 kpc, albeit with significant caveats as described in Lupu et al. At the same time our PDR region size is significantly smaller than the intrinsic source size obtained from our lensing model. This disagreement is not a big concern however, and may even be expected. Because the lensing model is based on a 1.5 micron image, i.e. ~397 nm rest frame, it is sensitive to the total stellar component in SDP11. Our PDR analysis is sensitive only to the massive and young O and B stars necessary for the ionization and excitation of [CII]. As such we expect the PDR derived size to be smaller than our lensing model. This also implies that we cannot use our PDR source size to better constrain the results of our lensing model.

4. Discussion

Our analysis suggests that SDP11 is a highly lensed analogue of a local ULIRG system in the containing a compact star formation region and concomitant high FUV fields, albeit in the Early Universe. Unfortunately, the true intrinsic source luminosity depends on our adopted magnification factor, which is not well constrained. A useful diagnostic plot that can help confirm our conclusion on the nature of SDP11 is the plot of $L_{[CII]}/L_{FIR}$ versus $L_{CO(1-0)}/L_{FIR}$ as described in Stacey et al. 2010 and shown in Figure 5. Since in the star forming paradigm these lines and the FIR continuum arise in the same regions, these ratios should be insensitive to variations in magnification caused by slight difference in location of the source relative to the critical curve of the lens. In Figure 5 we identify typical regions occupied by nearby sources as well as FUV radiation and density contours from the PDR models of Kaufman et al. (2006). If a source falls in the lower-right section of the plot with [CII]/CO $\leq 4100$, then the observed flux to continuum ratios can be explained fully in a star-forming paradigm. Sources that fall in the upper-left half of the plot may have significant [CII] emission from non-PDR sources such as XDRs within AGN (Stacey et al. 2010), low density ionized gas, or low metallicity molecular clouds (Stacey et al. 1991).

The lowest CO rotational transition detected from SDP11 is the CO 2-1 line from Riechers et al. (2013, in prep). Based on the PDR model solution from Section 6 above we can estimate the ratio of CO(2-1) to CO(1-0) and hence the strength of the CO(1-0) line. For our model derived FUV field strength and gas density the expected CO(2-1)/CO(1-0) line-integrated flux ratio is 7:1, agreeing with observations of both local and high-z systems (Downes & Solomon 1998; Bradford et al. 2003; Weiss et al. 2005b; Ward et al. 2003; Israel 2009; Weiss et al. 2005a; Aravena et al. 2008). Using this estimate along with the observed [CII] and FIR luminosity we place SDP11 on the plot of $L_{[CII]}/L_{FIR}$ versus $L_{CO(1-0)}/L_{FIR}$. The location of SDP11 clearly places it in the ULIRG region of the plot. More generally, because it falls in the allowed region of the plot, we have further support that our analysis in the PDR paradigm and PDR derived source size are indeed correct. As such SDP 11 does indeed appear analogous to local ULIRGs in terms of luminosity, FUV field strength, and source size, and may contains a starburst produced through the interaction between two Milky Way sized galaxies.

To further confirm the nature of SDP11 we can ask where it falls in the gas mass–star formation relation. Genzel et al. 2010 study the relations between the star formation rate and the molecular gas in galaxies from z ~ 0 to ~3. They find that quiescent star-forming galaxies at all epochs follow a similar relationship between their FIR luminosity and their CO luminosity, tracers of star formation and molecular gas respectively. However, merging systems—both local ULIRGs and high-z mergers—lie above the sequence of quiescently star-forming galaxies suggesting a higher star formation efficiency in merging systems. Where is SDP11? The FIR luminosity and CO(2-1) luminosity scaled to the expected CO(1-0) luminosity as above, places SDP11 among the merging systems suggesting a merger-induced star formation efficiency for SDP11 higher than "normal" galaxies. Over the range of allowed magnification factors, SDP11 falls among the samples of local ULIRGS ($\mu$ =29) or high-z mergers ($\mu$ = 7) in the $L_{FIR}$–$L_{CO}$ plane of Genzel et al. (2010). Both of these samples lie on a $L_{FIR}$–$L_{CO}$ relation that is ~ 4 times that of normal galaxies.

Studies over the past decade have shown ULIRGs in the early universe to be a very diverse population. Evidence for different modes powering high-z ULIRGs, including AGN, mergers, and the accretion of gas from the IGM, have all been observed. The discovery of the latter mode, however, came as a great surprise. (c.f. Biggs & Ivison 2008; Tacconi et al. 2010;

Iono et al. 2009, Hailey-Dunsheath et al. 2010; Stacey et al. 2010). In local ULIRGs star formation is triggered by mergers of massive galaxies (Sanders & Mirabel 1996), and it was long thought that only an AGN or similar major-merger event could produce of the extreme luminosities in high-z systems. We now know the star formation process in some $z \sim 1 - 3$ systems can also be stimulated through the accretion of cold gas *(*Tacconi et al. 2012, Genzel et al. 2010*)*. The large-scale of these starbursts are best understood as Schmidt-Kennicutt law star-formation, with star-forming efficiencies similar to local "normal" galaxies, but with starburst-like star formation rates arising from the large molecular-gas reservoirs accreted from the cosmic web. Studies revealing the accretion mode in high-z galaxies have created a paradigm shift in our thinking about galaxy evolution and formation in the early universe. Recent models have even demonstrated that cold accretion and typical Kennicutt-type star formation efficiencies can fully account for observed black hole growth and stellar mass assembly in the early universe (Di Matteo et al. 2012, Kenes et al. 2005, Genel et al. 2012, Agetz et al. 2010, Kitzbicher & White 2007).

This is not so say that the mergers are uncommon or unimportant in the early universe. For one, observational and theoretical evidence suggests a higher merger rate in the past (Lotz et al. 2011, Hopkins et al. 2010, Bertone & Conselice 2009). Furthermore, the relative importance and prevalence of merger driven versus accretion driven star formation in the early universe is still highly contested due to observational challenges in identifying merger events at high-z as well as measuring the full extent and size of gas disks. With the advent of high spatial-resolution submillimeter/millimeter observatories (e.g. PdBI, SMA, and ALMA) we are now just beginning to adequately address this question.

Recently Bothwell et al. (2013) examine the CO and FIR emission of 40 luminous ($L_{FIR} \gtrsim 10^{12}\, L_\odot$) galaxies at z ~ 1.2 – 4.1 using the PdBI and selected for in the submillimeter (i.e. submillimeter galaxies, SMG). The authors conclude that 20 – 28% of their sources show signs of an ongoing merger, yet within the observational constraints there is no evidence suggesting increased star formation efficiencies as one might expect for a merger and as seen in works like Genzel et al. (2010) mentioned above. Another indicator of the star formation efficiency in a galaxy is where it falls in the SFR and stellar mass relation (SFR–$M_*$) which has shown a tight relation between the SFR and stellar-mass in normal galaxies (i.e. accretion mode) with them falling along a "main sequence" (Brinchmann et al. 2004; Noeske et al. 2007; Elbaz et al. 2007). Merging systems lie above the main sequence (Elbaz et al. 2011). Hung et al. (2013) examine the SFR–$M_*$ relation and morphology of the 0.2 < z < 1.5 Herschel-selected galaxies. Based on morphological classification Hung et al. find that the fraction classified as "irregular" (indicating a merger) increases with IR luminosity at all values of $M_*$. Furthermore, they find that the fraction of interacting or merging systems increases with $L_{IR}$ as well as distance above the main sequence, with nearly 50% of galaxies showing evidence of a merger at $L_{IR} > 10^{11.5}\, L_\odot$. Perhaps most interestingly however is Hung et al.'s finding that ≳ 18% of IR-luminous galaxies on the main-sequence show evidence of interactions and mergers suggesting that evolution of galaxies on the MS may not be explained by gas accretion and exhaustion alone.

Clearly there is much work yet to be done in understanding the various modes of star formation and their impact in the early universe. For instance one might ask, what is the product of a major merger at early times? One possibility is that mergers, while not responsible for the bulk of star formation in the universe, are responsible for the formation of modern-day giant ellipticals. Quite recently Fu et al. (2013) report on the discovery of a system, HXMM01, featuring two submillimeter galaxies in the early stages of a merger. Based on the gas depletion time scales and star formation rates these authors conclude that HXMM01 may indeed be the progenitor of a modern day elliptical. The merging galaxies in HXMM01 feature star-forming regions of ~1.4 kpc in diameter, similar to the size of the star forming regions we estimate for SDP11. Perhaps then SDP11 represents a later stage in the merger process than HXMM01, in which the galaxies have already coalesced.

To test the various models of galaxy formation and constrain galaxy evolution scenarios we need to look for signs of different modes of star formation, i.e. accretion versus merger driven. Yet, identifying merging systems has proven to be a challenge at high-z. Morphological determinations like those in Hung et al. (2013) require both high sensitivity and high spatial resolutions. Until recently this was easiest to do at visible wavelengths, which of course is compromised by the large dust content of many IR luminous systems at high-z. Now however ALMA allows us to probe the morphologies of early galaxies in their rest frame IR emission, like the work in Fu et al. (2013) mentioned above, reducing the challenges of morphological classification. Using ZEUS-2 we can perform a census of the high-z analogues to local ULIRGs by observing their fine structure line emission. By combining ZEUS-2 and ALMA observations we can morphologically calibrate various FIR fine-structure line ratios in high-z systems to either merger or accretion driven modes of star formation instead of relying on extrapolations from local systems. This will allow us to classify the mode of star formation in systems that are spatially unresolved by ALMA. Lastly, in addition to assessing their total numbers, we can hope to identify the numbers of high-z ULIRG analogues in their various stages of merger to provide detailed constraints on the peak of the major merger rate in the Universe and the effects of mergers on galaxy properties.

5. Summary

We have built a new submillimeter spectrometer, ZEUS-2, and recently commissioned it on the APEX telescope detecting the [CII] 158 micron line from SDP11 at z~1.8. Combining our ZEUS-2 observations with a new Herschel detection of the [OI] 63 micron line, archival HST images, and data from the literature we determine the following:

1) SDP11 is strongly gravitationally lensed with magnification factor $\mu$ ~ 7—29;

2) an analysis of the [CII], [OI], and FIR emission within a PDR paradigm yields FUV fields of G~20,000 $G_0$, gas densities of ~2300 $cm^{-3}$, and a source size between 0.4 and 0.9 kpc in diameter for SDP11, all of which are similar to values seen in local ULIRGs;

3) and we conclude that SDP11 is likely a high-z analog of local ULIRGs featuring a compact and intense starburst confined to sub-kiloparsec scales and induced by a merger of two or more galaxies.

This work demonstrates the science that ZEUS-2 is capable of providing. Due to its synergy with ALMA and it being the only spectrometer capable of observing between 200 and 300 microns in the post Herschel era, we expect ZEUS-2 will be an important tool for studying galaxies in the years to come.

Acknowledgements

Part of this work is based on observations made with the NASA/ESA Hubble Space Telescope, and obtained from the Hubble Legacy Archive, which is collaboration between the Space Telescope Science Institute (STScI/NASA), the Space Telescope European Coordinating Facility (ST-ECF/ESA) and the Canadian Astronomy Data Centre (CADC/NRC/CSA). ZEUS-2 development and observations are supported by NSF grants AST-0705256, AST-0722220,AST-1105874, and AST-1109476, NASA grant NNX10AM09H, and a grant from Georgia Southern University. C. F. would like to thank R. Wang for sharing her list of high-z [CII] detections from the literature and in Wang et al. 2013. Lastly the authors would like thank the anonymous referee for their helpful comments and acknowledge the APEX staff whose excellent support helped to make this work possible.

References:

Agertz, O., Teyssier, R. & Moore, B. 2011, *MNRAS*, 410, 1391–1408

Aravena, M., Bertoldi, F., Schinnerer, E., *et al.* 2008, *A&A*, 491, 173–181

Bertone, S. & Conselice, C. J. 2009, *MNRAS*, 396, 2345–2358

Biggs, A. D. & Ivison, R. J. 2008, *MNRAS*, 385, 893–904

Bothwell et al. 2013, MNRAS, 429, 3047-3067

Bradford, C. M., Nikola, T., Stacey, G. J., *et al.* 2003, *ApJ*, 586, 891–901

Brauher, J. R., Dale, D. A. & Helou, G. 2008, *ApJS*, 178, 280–301

Brinchmann et al. 2004, MNRAS, 351, 1151-1179

Brisbin, et al. 2013, *ApJ*, in prep.,

Carilli, C. L., Riechers, D., Walter, F., *et al.* 2013, *ApJ*, 763, 120

Colbert, J. W., Malkan, M. A., Clegg, P. E., *et al.* 1999, *ApJ*, 511, 721–729

Coppin, K. E. K., Danielson, A. L. R., Geach, J. E., *et al.* 2012, *MNRAS*, 427, 520–532

Coppin, K., Chapin, E. L., Mortier, A. M. J., *et al.* 2006, *MNRAS*, 372, 1621–1652

Cox, P., Krips, M., Neri, R., *et al.* 2011, *ApJ*, 740, 63

Crawford, M. K., Lugten, J. B., Fitelson, W., Genzel, R. & Melnick, G. 1986, *ApJ*, 303, L57

Daddi, E., Dickinson, M., Morrison, G., *et al.* 2007, *ApJ*, 670, 156–172

De Breuck, C., Maiolino, R., Caselli, P., *et al.* 2011, *A&A*, 530, L8

Decarli, R., Walter, F., Neri, R., *et al.* 2012, *ApJ*, 752, 2

Di Matteo, T., Khandai, N., DeGraf, C., *et al.* 2012, *ApJ*, 745, L29

Downes, D. & Solomon, P. M. 1998, *ApJ*, 507, 615–654

Eales, S., Dunne, L., Clements, D., *et al.* 2010, *PASP*, 122, 499–515

Elbaz, D., Dickinson, M., Hwang, H. S., *et al.* 2011, *A&A*, 533, A119

Elbaz et al. 2007, A&A, 468, 33-48

Ferkinhoff, C., Hailey-Dunsheath, S., Nikola, T., *et al.* 2010, *ApJ*, 714, L147–L151

Ferkinhoff, C., Brisbin, D., Nikola, T., *et al.* 2011, *ApJ*, 740, L29

Fu, H., Cooray, A., Feruglio, C., *et al.* 2013, *Nature*, advance on,

Gallerani, S., Neri, R., Maiolino, R., *et al.* 2012, *A&A*, 543, A114

Genel, S., Naab, T., Genzel, R., *et al.* 2012, *ApJ*, 745, 11

Genzel, R., Tacconi, L. J., Gracia-Carpio, J., *et al.* 2010, *MNRAS*, 407, 2091–2108

Graciá-Carpio, J., Sturm, E., Hailey-Dunsheath, S., *et al.* 2011, *ApJ*, 728, L7

Hailey-Dunsheath, S., Nikola, T., Stacey, G. J., *et al.* 2010, *ApJ*, 714, L162–L166

Hailey-Dunsheath, S., et al. 2013, *ApJ*, in prep.

Helou, G., Soifer, B. T. & Rowan-Robinson, M. 1985, *ApJ*, 298, L7

Hildebrand, R. H., Loewenstein, R. F., Harper, D. A., *et al.* 1985, *Icarus*, 64, 64–87

Hopkins, P. F., Croton, D., Bundy, K., *et al.* 2010, *ApJ*, 724, 915–945

Hung et al. 2013, Accepted in ApJ, arXiv: 1309.4459

Iono, D., Wilson, C. D., Yun, M. S., *et al.* 2009, *ApJ*, 695, 1537–1549

Iono, D., Yun, M. S., Elvis, M., *et al.* 2006, *ApJ*, 645, L97–L100

Israel, F. P. 2009, *A&A*, 493, 525–538

Ivison, R. J., Papadopoulos, P. P., Smail, I., *et al.* 2011, *MNRAS*, 412, 1913–1925

Ivison, R. J., Swinbank, A. M., Swinyard, B., *et al.* 2010, *A&A*, 518, L35

Joy, M., Lester, D. F. & Harvey, P. M. 1987, *ApJ*, 319, 314

Kaufman, M. J., Wolfire, M. G., Hollenbach, D. J. & Luhman, M. L. 1999, *ApJ*, 527, 795–813

Kaufman, M. J., Wolfire, M. G. & Hollenbach, D. J. 2006, *ApJ*, 644, 283–299

Kaufman, M. J., Wolfire, M. G. & Hollenbach, D. J. 2006, *ApJ*, 644, 283–299

Keeton, C. R. 2001, 17

Keres, D., Katz, N., Weinberg, D. H. & Dave, R. 2005, *MNRAS*, 363, 2–28

Kitzbichler, M. G. & White, S. D. M. 2007, *MNRAS*, 376, 2–12

Lord, S. D., Hollenbach, D. J., Haas, M. R., *et al.* 1996, *ApJ*, 465, 703

Lotz, J. M., Jonsson, P., Cox, T. J., *et al.* 2011, *ApJ*, 742, 103

Luhman, M. L., Satyapal, S., Fischer, J., *et al.* 1998, *ApJ*, 504, L11–L15

Luhman, M., Satyapal, S., Fischer, J., *et al.* 2003, *ApJ*, 594, 758

Lupu, R. E., Scott, K. S., Aguirre, J. E., *et al.* 2012, *ApJ*, 757, 135

Maiolino, R., Caselli, P., Nagao, T., *et al.* 2009, *A&A*, 500, L1–L4

Maiolino, R., Cox, P., Caselli, P., *et al.* 2005, *A&A*, 440, L51–L54

Maiolino, R., Gallerani, S., Neri, R., *et al.* 2012, *MNRAS: Letters*, 425, L66–L70

Malhotra, S., Kaufman, M. J., Hollenbach, D., *et al.* 2001, *ApJ*, 561, 766–786

Marsden, G., Borys, C., Chapman, S. C., Halpern, M. & Scott, D. 2005, *MNRAS*, 359, 43–46

Meijerink, R., Spaans, M. & Israel, F. P. 2006, *ApJ*, 650, L103–L106

Nagao, T., Maiolino, R., De Breuck, C., *et al.* 2012, *A&A*, 542, L34

Negishi, T., Onaka, T., Chan, K.-W. & Roellig, T. L. 2001, *A&A*, 375, 566–578

Negrello, M., Hopwood, R., De Zotti, G., *et al.* 2010, *Science (New York, N.Y.)*, 330, 800–4

Noeske et al. 2007

Oberst, T. E., Parshley, S. C., Stacey, G. J., *et al.* 2006, *ApJ*, 652, L125–L128

Pound, M. W. & Wolfire, M. G. 2008, *Astronomical Data Analysis Software and Systems ASP Conference Series*, 394,

Rice, W., Lonsdale, C. J., Soifer, B. T., *et al.* 1988, *ApJS*, 68, 91

Riechers, D. A., Bradford, C. M., Clements, D. L., *et al.* 2013, *Nature*, 496, 329–33

Sanders, D. B. 2000, *Advances in Space Research*, 25, 2251–2264

Sargsyan, L., Lebouteiller, V., Weedman, D., *et al.* 2012, *ApJ*, 755, 171

Siebenmorgen, R. & Krügel, E. 2007, *A&A*, 461, 445–453

Smail, I., Ivison, R. J. & Blain, A. W. 1997, *ApJ*, 490, L5–L8

Spergel, D. N., Verde, L., Peiris, H. V., *et al.* 2003, *ApJS*, 148, 175–194

Stacey, G. J., Geis, N., Genzel, R., *et al.* 1991, *ApJ*, 373, 423

Stacey, G. J., Hailey-Dunsheath, S., Ferkinhoff, C., *et al.* 2010, *ApJ*, 724, 957–974

Sturm, E., Verma, A., Graciá-Carpio, J., *et al.* 2010, *A&A*, 518, L36

Swinbank, A. M., Karim, A., Smail, I., *et al.* 2012, *MNRAS*, 427, 1066–1074

Tacconi, L. J., Genzel, R., Neri, R., *et al.* 2010, *Nature*, 463, 781–4

Tacconi, L. J., Genzel, R., Smail, I., *et al.* 2008, *ApJ*, 680, 246–262

Tacconi, L. J., Neri, R., Chapman, S. C., *et al.* 2006, *ApJ*, 640, 228–240

Tacconi, L. J., Neri, R., Genzel, R., *et al.* 2013, *ApJ*, 768, 74

Tacconi, L., Neri, R. & Genzel, R. 2012, *arXiv preprint,* 1211.5743, 1–63

Tielens, A. G. G. M. & Hollenbach, D. 1985, *ApJ*, 291, 722

Valtchanov, I., Virdee, J., Ivison, R. J., *et al.* 2011, *MNRAS*, 415, 3473–3484

Venemans, B. P., McMahon, R. G., Walter, F., *et al.* 2012, *ApJ*, 751, L25

Wagg, J., Wiklind, T., Carilli, C. L., *et al.* 2012, *ApJ*, 752, L30

Wang, R., Wagg, J., Carilli, C. L., *et al.* 2013, 24

Ward, J. S., Zmuidzinas, J., Harris, A. I. & Isaak, K. G. 2003, *ApJ*, 587, 171–185

Weiß, A., Downes, D., Walter, F. & Henkel, C. 2005, *A&A*, 440, L45–L49

Weiß, A., Kovács, A., Coppin, K., *et al.* 2009, *ApJ*, 707, 1201–1216

Weiß, A., Walter, F. & Scoville, N. Z. 2005, *A&A*, 438, 533–544

Wilkins, S. M., Matteo, T. D., Croft, R., *et al.* 2013, *MNRAS*, 429, 2098–2103

Wilson, C. D., Harris, W. E., Longden, R. & Scoville, N. Z. 2006, *ApJ*, 641, 763–772

Wolfire, M. G., Tielens, A. G. G. M. & Hollenbach, D. 1990, *ApJ*, 358, 116

Table 1: H-ATLAS J091043.1-000322 Source Parameters

| Parameter | Unit | Value | Source |
|---|---|---|---|
| RA | hh:mm:ss.s | 9:10:43.1 | Lupu et al. 2012 |
| DEC | dd:mm:ss | -00:03:22 | " |
| z | ... | 1.786 | Negrello et al. 2010 |
| $\mu$ | ... | 18 ± 11 | this work |
| $L_{FIR}$ | $\mu \cdot 10^{13} L_\odot$ | 3.88 ± 0.47 | this work |
| $M_{H2}$ | $\mu \cdot 10^{11} M_\odot$ | 2.0 | Lupu et al. 2012 |
| $\mu \cdot \Omega_d$ | arcsec$^2$ | 0.43 | Lupu et al. 2012 |
| F([CII] 158) | $\mu \cdot 10^{-18}$ W m$^{-2}$ | 6.44 ± 0.42 | this work |
| F([OI] 63) | " | 7.5 ± 2.3 | Verma et al., in prep. |
| CO(2-1) | " | 0.0234 ± 0.00257 | Riechers et al., in prep |
| CO(5-4) | " | 0.159 ± 0.053 | Lupu et al. 2012 |
| CO(6-4) | " | 0.241 ± 0.083 | " |
| CO(7-6) | " | 0.174 ± 0.136 | " |
| [CI] 370 | " | 0.301 ± 0.136 | " |

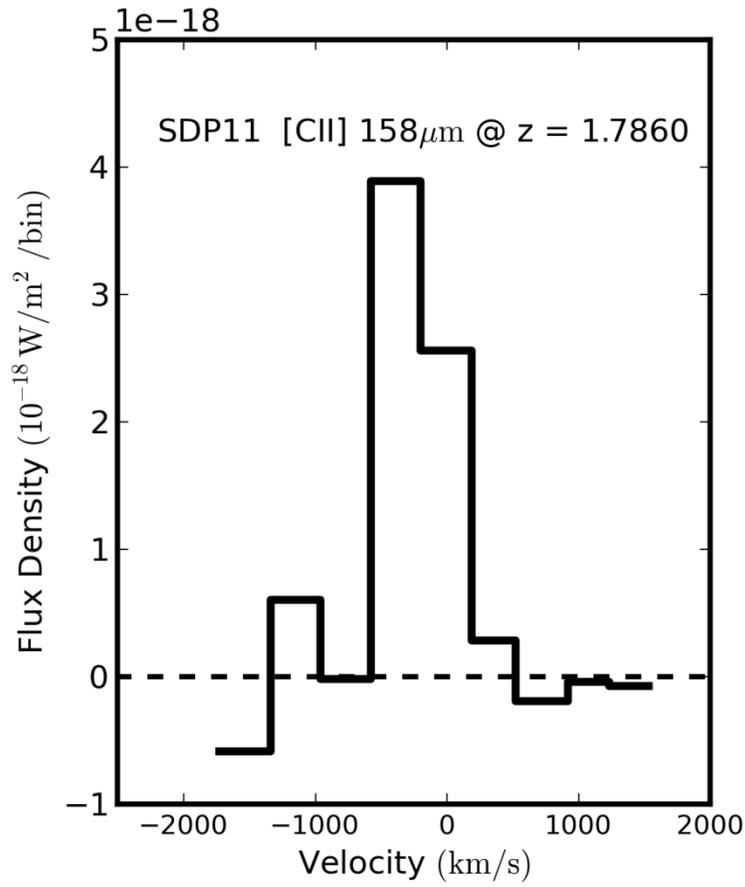

Figure 1: ZEUS-2/APEX detection of the [CII] 158 micron line from H-ATLAS J091043.1-000322 plotted versus the source rest-frame velocity. Spectral bins are ~ 1 resolution element and equal to ~350 km/s. The continuum emission has been removed.

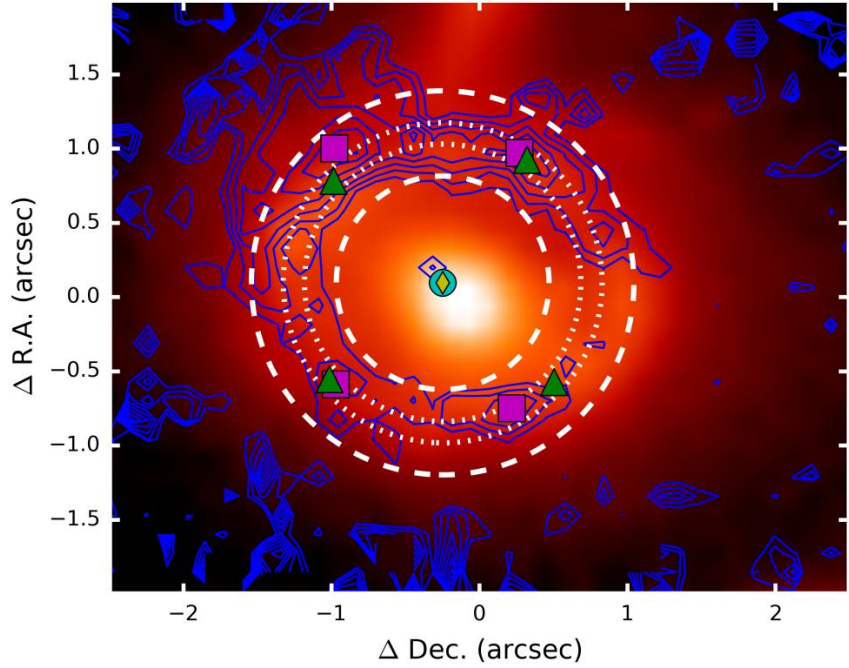

Figure 2: HST/WFC3 F110W image with F160W/ F110W divided image overlaid in blue contours. The squares denote the positions of the peaks in the divided image used for the gravitational lens model described in Section 2.3. The triangles show the best-fit positions of the emission peaks while the diamond locates the best-fit lens and source positions. The white contours are the 70% of the peak-flux limits for Einstein rings produced by our best-fit lens model and an extended source with Gaussian light profile and half-light diameters of 0.7 kpc (dotted) and 3.4 kpc (dashed).

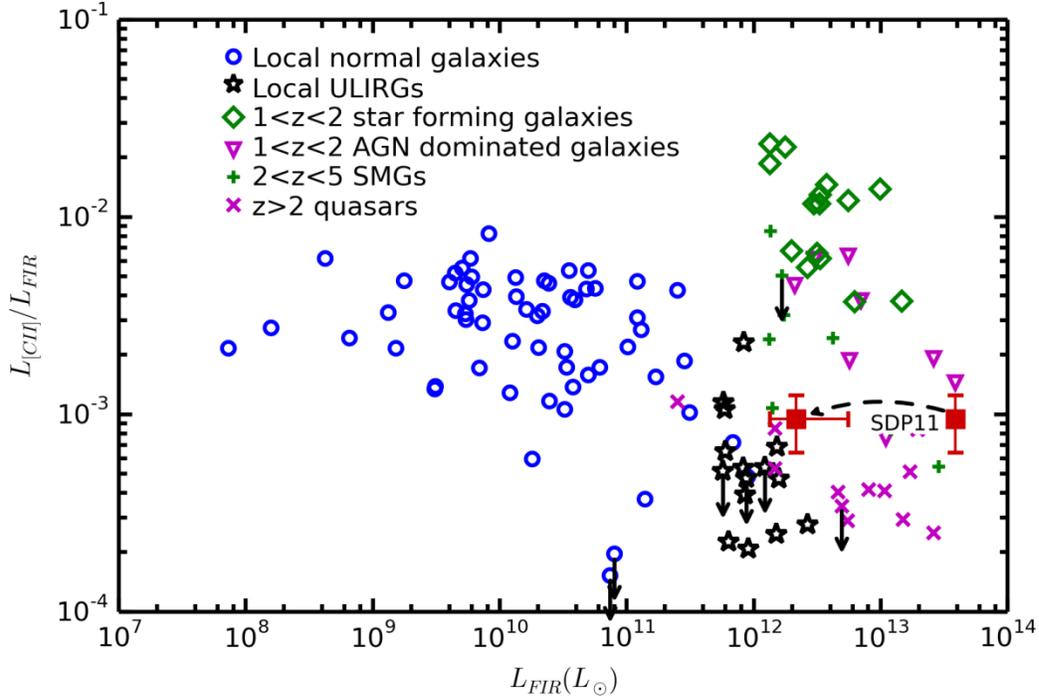

Figure 3: $L_{[CII]}/L_{FIR}$ ratio as a function of the $L_{FIR}$ for local and high redshift galaxies. The ratio from SDP11 (square) is plotted versus both its apparent and intrinsic FIR luminosity. The effect of delensing SDP11 is noted with an arrow while the horizontal error bars indicated the effects of the allowed magnification factors. For comparison we have also included ratios for local normal star forming galaxies (Malhotra et al. 2001), local ULIRGs (Luhman et al. 2003), the ZEUS-1 star forming and AGN samples (Stacey et al. 2010, Brisbing et al., in prep.; Hailey-Dunsheath et al., in prep.), z > 2 submillimeter bright galaxies (SMG; Maiolino et al. 2009; Ivison et al. 2010; De Breuck et al. 2011; Swinbank et al. 2012; Wagg et al. 2012; Valtchanov et al. 2011; Riechers et al. 2013), and z>2 quasars (Pety et al. 2004; Maiolino et al. 2005, 2009; Gallerani et al. 2012; Wagg et al. 2012; Carilli et al. 2013; Maiolino et al. 2005; Leipski et al. 2013; Willott et al. 2013; Venemans et al. 2012)

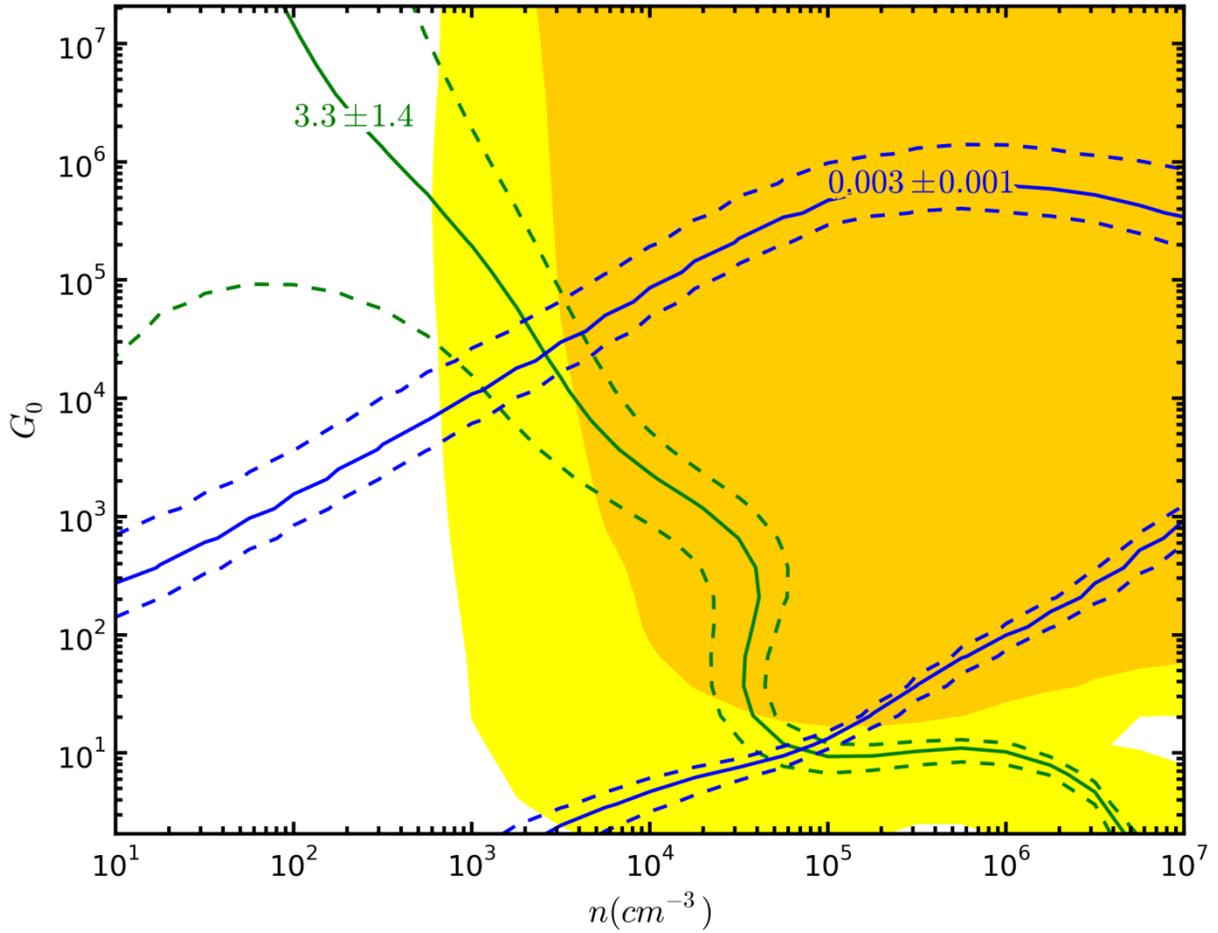

Figure 4: PDR modeling using the online PDR Toolkit (Pound & Woolfire 2008, Kaufman et al 2006) of the corrected [OI]/[CII] ratio (green), ([OI]+[CII])/FIR ratio (blue) from SDP11. The shaded regions show the allowed $G_0$ and $n$ phase space of the mid-J CO observations of Lupu et al. 2012 where yellow is the ratio of [CI] 370 µm + CO(7-6) to CO(5-4) and orange is the CO(6-5)/CO(5-4) ratio. The dash lines indicate ±1σ error bounds of the ratios listed above. For all lines we have corrected for ionized gas and optical depth effects as described in the text.

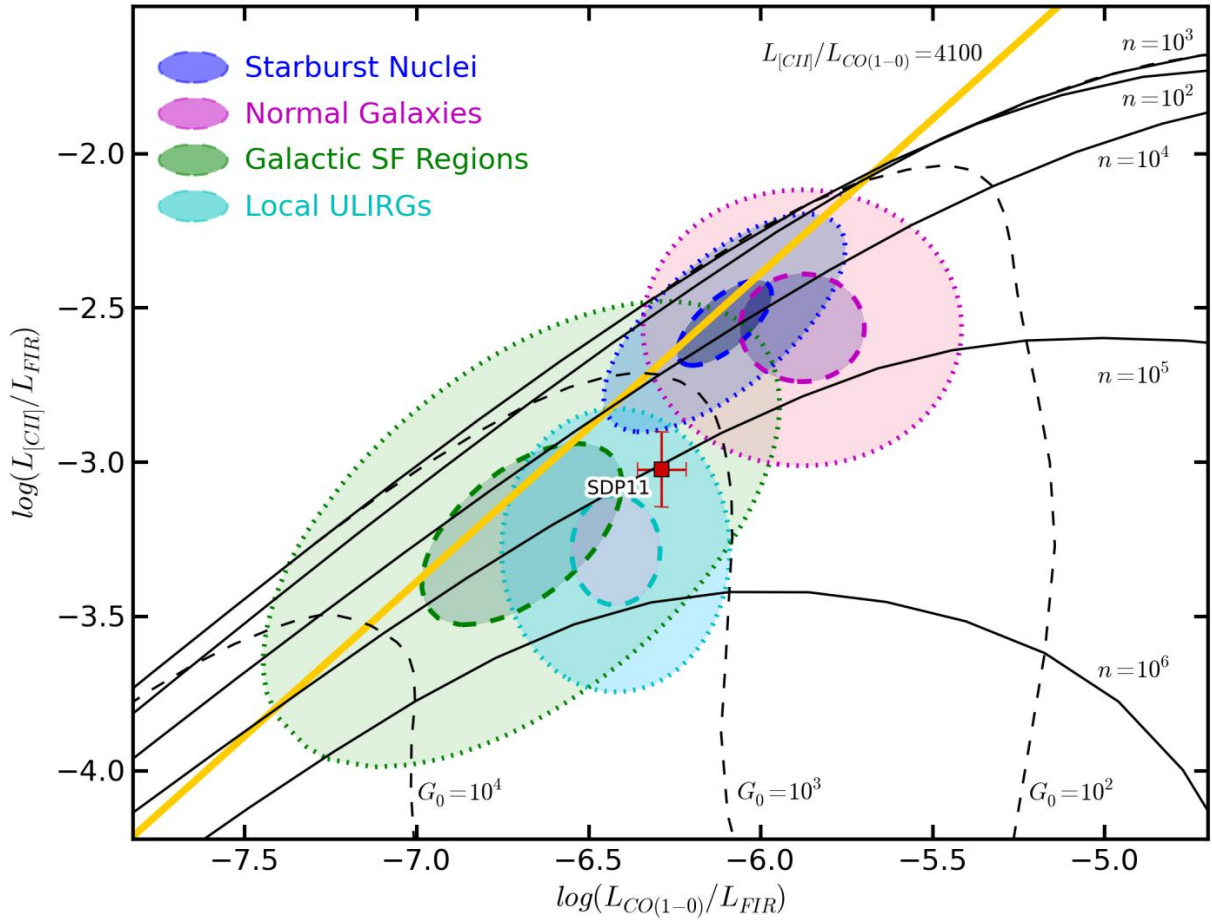

Figure 5: The log(L$_{[CII]}$/L$_{FIR}$ ) versus log(L$_{CO(1-0)}$/L$_{FIR}$) for SDP11 adapted from Stacey et al. 2010. Overlaid are the FUV radiation field and gas density contours from Kauffmann et al 2006. The typical L$_{[CII]}$/L$_{CO(1-0)}$ ratio for star-forming galaxies (~4100) is shown by the thick line. Regions occupied by galactic star forming regions, starburst galaxies, normal galaxies and local ULIRGs are illustrated with 90% (dashed) and 50% (dotted) maximum contours of 2D Gaussians fit to the various source samples from literature (e.g. Stacey et al. 1991, Malhotra et al. 2001, Luhman et al. 2003)